\documentstyle[prb,aps,twocolumn,psfig]{revtex}
\begin{document}
\draft
\title{Cohesive properties of alkali halides}
\author{
  Klaus Doll}
\address{
     Max-Planck-Institut f\"ur Physik komplexer Systeme,
          Bayreuther Str.~40, D-01187 Dresden, Germany }
\author{Hermann Stoll}

\address{
Institut f\"ur Theoretische Chemie, Universit\"at Stuttgart, D-70550 Stuttgart,
Germany
}

\maketitle

\begin{abstract}
We calculate cohesive properties of LiF, NaF, KF, LiCl, NaCl, and KCl 
with {\em ab-initio} quantum chemical methods. 
The coupled-cluster approach is used to correct the Hartree-Fock 
crystal results
for correlations and to systematically improve
cohesive energies, lattice constants
and bulk moduli. After inclusion of correlations, we recover 95-98\% of the
total cohesive energies. The lattice constants deviate from experiment 
by at most 1.1\%, bulk moduli
by at most 8\%. We also find good agreement for spectroscopic properties
of the corresponding diatomic molecules.
\end{abstract}

\pacs{ }

\narrowtext
\section{Introduction}

One of the
earliest methods for a quantitative treatment of the cohesion of ionic solids 
was the Born-Mayer theory \cite{Tosi}. L\"owdin \cite{Loewdin}
made a first quantum-mechanical approach starting from the symmetrically
orthogonalized orbitals of the free ions; these orbitals were used to 
approximate the density matrix and to calculate
the Hartree-Fock energy. Since the advent of
density-functional theory and especially the local density approximation, 
the latter methods have become standards in solid state
physics \cite{JonesGunnarsson}. 
However, there has also been progress in the development of 
wavefunction based 
methods. Hartree-Fock (HF) calculations can be done routinely nowadays with
the help of the program package {\sc Crystal} \cite{CRYSTAL}, and it is even
possible to include electron correlations.
One way of achieving that is by multiplying the HF wavefunction
with a Jastrow factor containing several parameters; these parameters can
be optimized with the help of Monte-Carlo methods \cite{QMC}.
A first attempt to include 
correlations by means of quantum chemical methods was made using
the Local Ansatz \cite{Stollhoff,Fuldebuch};
here local excitation operators are applied for modifying the HF wavefunction.
In the last years, an "incremental scheme" (an expansion of the total
correlation energy in terms of one-body, two-body, three-body and higher
contributions, the so called "local increments")
has been
developed and successfully applied to semiconductors \cite{SEMIC}. This  
method has been extended to ionic solids and applied to several oxides 
(MgO, CaO, NiO)\cite{DDFS}.
Alkali halides are model examples of ionic solids and have recently
been carefully investigated at the HF level \cite{Prencipe}. 
The major part of the experimental
lattice energy is already recovered at this level.
However, the lattice constants significantly deviate from the experimental
values, especially for the heavier compounds. 
We want to show that the incremental scheme 
can explain the deviations of the HF results from experiment. 

\section{The Method}
\subsection{Incremental scheme}
The scheme has been explained in earlier work and we only repeat
the main ideas. The correlation energy of the solid is expanded into a sum
of local contributions (increments)
\begin{eqnarray*}
\epsilon_{\rm bulk} = \sum_A \epsilon(A) + \frac{1}{2} \sum_{A,B}
\Delta\epsilon(AB) \\ + \frac{1}{3!} \sum_{A,B,C}
\Delta\epsilon(ABC) + ...
\end{eqnarray*}
$\epsilon(A)$ is the correlation energy of a  group of localized
orbitals (a so-called one-body increment), the non-additivity 
$\Delta \epsilon(AB)=\epsilon(AB)-\epsilon(A)-\epsilon(B) \mbox{ }$
defining a two-body increment, and so on.
Usually, this series is evaluated up to three-body increments.
The increments are extracted from clusters containing up to three
explicitly described ions (i.e. ions with a high-quality basis set) embedded
in a set of pseudopotentials and point charges. They should be 
well transferable, which means that they should only weakly depend on the
specific cluster chosen for their evaluation 
(e.g. the value of a one-body increment obtained from
a cluster with one explicitly described ion only weakly
varies when extracted from a cluster with more than one explicitly 
described ion).
As correlation scheme, we chose the coupled-cluster approach
with single and double substitutions (CCSD)\cite{CCSD} with
an exponential ansatz for the correlated wavefunction:
\begin{eqnarray*}
|\Psi_{\rm CCSD}>=\mbox{exp}\left({\sum_{a \atop r}c_a^r a^+_r a_a +
\sum_{a<b \atop r<s}c_{ab}^{rs}a^+_r a^+_s a_a a_b}\right)|\Psi_{\rm SCF}>.
\end{eqnarray*}
In addition, we applied the CCSD(T)\cite{Raghavachari} scheme 
including triple excitations in a perturbative way.
All the calculations were done with the {\em ab-initio} program package 
{\sc Molpro}
\cite{KnowlesWerner,MOLPROpapers}.
Localization was done by the Foster-Boys method \cite{FosterBoys},
 and all of the $ns, np$ valence and outer-core orbitals of the halide
and alkali ions, respectively, ($n$ = 2 for F, Na and 3 for  Cl, K, $1s$ 
in the case of Li), were correlated. 

\subsection{Pseudopotentials and basis sets} 
The increments are taken from cluster calculations. The ions to be correlated
are accurately described with extended basis sets. Negatively charged ions
are embedded with X$^+$ pseudopotentials as next neighbors
to simulate the Pauli repulsion. 
Finally, the system is embedded in a set of point charges
(typically 7 $\times$ 7 $\times$ 7 
lattice sites with charges $\pm 1$ in the interior and reduced by factors
of 2, 4 and 8 at the surface planes, edges and corners, respectively
\cite{Evjen}). 
The description of the explicitly treated ions  is as follows.
We used a   $[5s4p3d2f]$ basis \cite{Dunning} for F  
and a $[6s5p3d2f]$ basis \cite{Dunning} for Cl. 
For Li, we used a $[5s4p3d2f]$ basis \cite{Dunning},
for Na a $[7s6p5d4f]$ basis (Ref. \onlinecite{Roos}, with $d$ and $f$ 
functions uncontracted). Finally, for 
K we used a 9-valence-electron pseudopotential \cite{Leininger} with 
the corresponding $sp$ basis set (uncontracted) and augmented with 
5 $d$ and 3 $f$ functions \cite{KBasis}, resulting in a
$[7s6p5d3f]$ basis.

\section{Results}
\subsection{Ionization potentials, electron affinities and results for
the diatomic molecules}

In Table \ref{EAIP}, we give results for atomic electron affinities and
ionization potentials. At the correlated level, we 
obtain good agreement with experiment (to $<$0.1 eV)
in all cases. Results for the
diatomic molecules are given in Table \ref{diatomic}. Again, we obtain nice 
agreement, to $\leq$ 0.02 \AA \mbox{} (1\%) for bond lengths,
24 cm$^{-1}$ (4\%) for vibrational frequencies,
and 0.1 eV for dissociation energies $D_e$.
Note that we calculated
$D_e$ as the difference $E_{atom 1}+E_{atom 2}-E_{diatomic}$,
in contrast to Ref. \onlinecite{Bauschlicher}, where the 
dissociation energy was first calculated 
with respect to the singly charged ions and
then corrected with the help of the experimental electron 
affinities and ionization potentials. The experimental dissociation energy
for NaF from Ref. \onlinecite{Huber-Herzberg} 
is probably too high, the experimental value
given in Ref. \onlinecite{Brewer} ($D_e$=4.97 eV) and the theoretical value
from Ref. \onlinecite{Bauschlicher} are closer to our calculated value.

\subsection{Results for the solid}

\subsubsection{Hartree-Fock calculations}
We repeated the {\sc{Crystal}} calculations from Ref. \onlinecite{Prencipe} 
with
essentially the same basis sets \cite{Kalium}. We calculated
both the lattice energy (cohesive energy with respect to the ions) as well
as the cohesive energy with respect to the neutral 
atoms.
The lattice energy
is already in good agreement with experiment. This is what one would expect 
since in 
purely ionic solids (the Mulliken population analysis gives 
a charge transfer  very close to $\pm$1 in all cases) the 
Madelung energy makes the most important contribution to the lattice energy;
the Madelung energy is already in rough agreement with experiment
\cite{AshcroftMermin}. However, the 
cohesive energy with respect to the atoms is less well described
as a consequence of the missing intra-atomic correlation effects.
Moreover, lattice constants are by up to $\sim$ 5\% too large at the HF level, 
bulk moduli up to $\sim$ 21\% too small.

\subsubsection{One-body increments}
Results for the crystal correlation energies 
are given in Tables 
\ref{alleInk},\ref{inkfluorides},\ref{inkchlorides},\ref{inkKCl657}.
Concerning the one-body increments, we obtain nearly the same correlation 
energy for the free alkali ions and the corresponding embedded ions. 
This is of course a consequence of the small ionic radii of the 
cations. In the case of the anions F$^-$ and Cl$^-$, 
we find that the absolute value of the
correlation energy in the solid is smaller than for the free ion, by up to 
0.4 eV. Such an effect was already found in the calculations on
the oxides \cite{DDFS} and is explained by the lower level
spacing of the excited states for the free ion compared to the embedded ion
where excitations are higher in energy.

\subsubsection{Two-body and three-body increments}
The two-body correlation-energy 
increments decrease rapidly. The decay is compatible with a
van der Waals law from second nearest neighbors on, 
cf. Table \ref{inkKCl657}.
By far the largest contributions come from next-neighbor 
metal-halide (M-X) and halide-halide (X-X) interactions.
The total effect of the M-X inter-atomic correlations is similar 
for X = F and X = Cl, but for given X increases
from Li to K (i.e.\ with increasing polarizability $\alpha$ of the metal ion) 
in such a way, that the ratio of the M-X contribution to the
X-X contribution changes from $<1$ to $>1$ (cf. Tables 
\ref{inkfluorides}, \ref{inkchlorides}). The X-X increments in turn are 
larger in magnitude for Cl than for F,
in agreement with the trend of the respective $\alpha$ values but in 
contrast to the situation for the intra-atomic difference in correlation
energies $\epsilon
({\rm free \mbox{ } ion})-\epsilon({\rm embedded \mbox{ } ion})$.
Quantitatively
comparing the F-F and Cl-Cl next-neighbor 
increments from different systems (Table 
\ref{vdWTabelle}) and assuming a purely van der Waals interaction,
we find that even in that case the van der Waals-law holds surprisingly
well. 
The $C_6$ coefficient can be determined from the two-body increments. For
the sake of simplicity, we assume a purely van der Waals interaction already
for next neighbors and for all types of correlations (e.g. also spin-flip
processes for Ni-O increments \cite{DDFS}).
The result for $C_6=\Delta E \times r^6$  obtained this way
is comparable to results
from literature, e.g. Refs. \onlinecite{Jain,Aquilanti} and references therein.

An estimate of the van der Waals interaction can be 
obtained using the London formula for dispersion interactions 
\cite{Atkins}: $E=-\frac{3}{2}\eta\frac{IP_1 IP_2}{IP_1 + IP_2}\frac{\alpha_1
\alpha_2}{r^6}$ with the ionization potentials (IP) as characteristic
excitation energies and polarizabilities
($\alpha$) of the two interacting systems ($\eta$ is of order unity, $r$ is
the distance). 
Polarizabilities and ionization potentials were calculated with the same
arrangement as the one-body increments: One ion with extended basis set
was embedded in  a set of point charges at the experimental lattice 
constant (and pseudopotentials as next 
neighbors, in the case of anions). To evaluate the polarizabilities, we 
applied a small dipolar field and find values in good agreement
with values from literature \cite{Fowler,Sangachin,Pyper}. 
The ionization potential was calculated with
the same cluster, which is certainly a crude approximation
because effects such as long-range polarization
are not included: the IP obtained this way is {\em not} 
what would be experimentally measured for the solid.
Our CCSD
results for the two-body increments 
are roughly 2 to 5 times larger (see Table \ref{vdWTabelle}) than
what is predicted from the London formula. This 
implies that the London formula can give a qualitative understanding of the
magnitude of the interionic interaction and the parameters describing it
($\alpha$, excitation energies), but is not able to predict
results quantitatively.
Van der Waals interactions in extended systems have also been considered
for He \cite{Staemmler} (see also a recent review \cite{Pyykkoe}). \\
We calculated three-body increments only for KCl (Tables \ref{inkchlorides} and
\ref{inkKCl657}). 
We find that
they are very small indicating a rapid convergence of the incremental 
expansion. Neglecting three-body
increments is not a serious approximation, therefore.

\subsubsection{Sum of increments and discussion}
The sums of the increments are given in Table \ref{alleInk}. 
Including correlations,
we obtain 95 to 98\% of the experimental cohesive energies. 
The relatively good agreement of the HF lattice energies already mentioned
above turns out to be due to 
a partial error cancellation. When the HF cohesive energies are calculated 
with respect to the free ions, the corrections due to the missing correlation 
effects have opposite signs: the one-body contributions diminish the 
cohesive energy since the absolute value for the free anion is higher than
that for the embedded ion; on the other hand, the van der Waals interactions
which are also missing at the HF level lead to an increase of the cohesive 
energy (cf.\ Tables \ref{inkfluorides}, \ref{inkchlorides}).
The compensation is nearly perfect for LiF, but already for KF the 
inter-atomic correlation effects overcompensate the intra-atomic
ones by nearly a factor of 2, and the weight is still further shifted 
in favour of the two-body effects for the MCl crystals,
so that for KCl, e.g., a factor of $\sim$6 is reached.

After inclusion of correlations, the lattice constants deviate by at most
1.1\% from experiment. As already found in the context of the oxides, the
one-body increments would enforce larger lattice constants (the absolute value
of the correlation 
energy of an anion increases when the lattice constant increases because
of the lower level spacing at larger lattice constant). The large reduction
of the lattice constants, on the other hand, is a two-body effect  resulting
from the van der Waals interaction between
the ions. 
The CCSD(T) results turn out to be slightly superior to CCSD \cite{CCSDT}.

At fixed lattice constant, inclusion of correlations leads to a 
decrease of the bulk modulus. However, for most of the solids considered here
correlations
reduce the lattice constant. This means that the HF bulk modulus has to be 
calculated at a smaller lattice constant where it increases again. As a net
result,
correlations increase the bulk moduli in most cases. 
Note that the bulk
moduli are  more sensitive to the fitting procedure than
cohesive energies and lattice constants and that they also 
have  large experimental uncertainties 
even at room temperature
(see the comparison in Ref. \onlinecite{Cortona}).

\begin{figure}
\caption{CCSD correlation-energies for KCl as a function 
of the lattice constant. The two-body
increments are already multiplied with the corresponding weight factors.
Displayed is the difference of correlation energy 
$\epsilon{\rm (embedded\mbox{ }Cl}^-)$ -$\epsilon{\rm (free\mbox{ } Cl}^-)$
(dashed - - -), the two-body increment 
Cl-Cl for next neighbors 
(dashed-dotted - $\cdot$ - ), the two-body increment K-Cl for next neighbors
(dotted $\cdot$ $\cdot$ $\cdot$), 
and the sum of these three correlation energies (solid 
line $\frac{ \mbox{ }\mbox{ }\mbox{ }\mbox{ } }{}$
) which make the most part of the total correlation contribution to the 
cohesive energy.}
\vspace{0.5cm}
\centerline{\psfig{figure=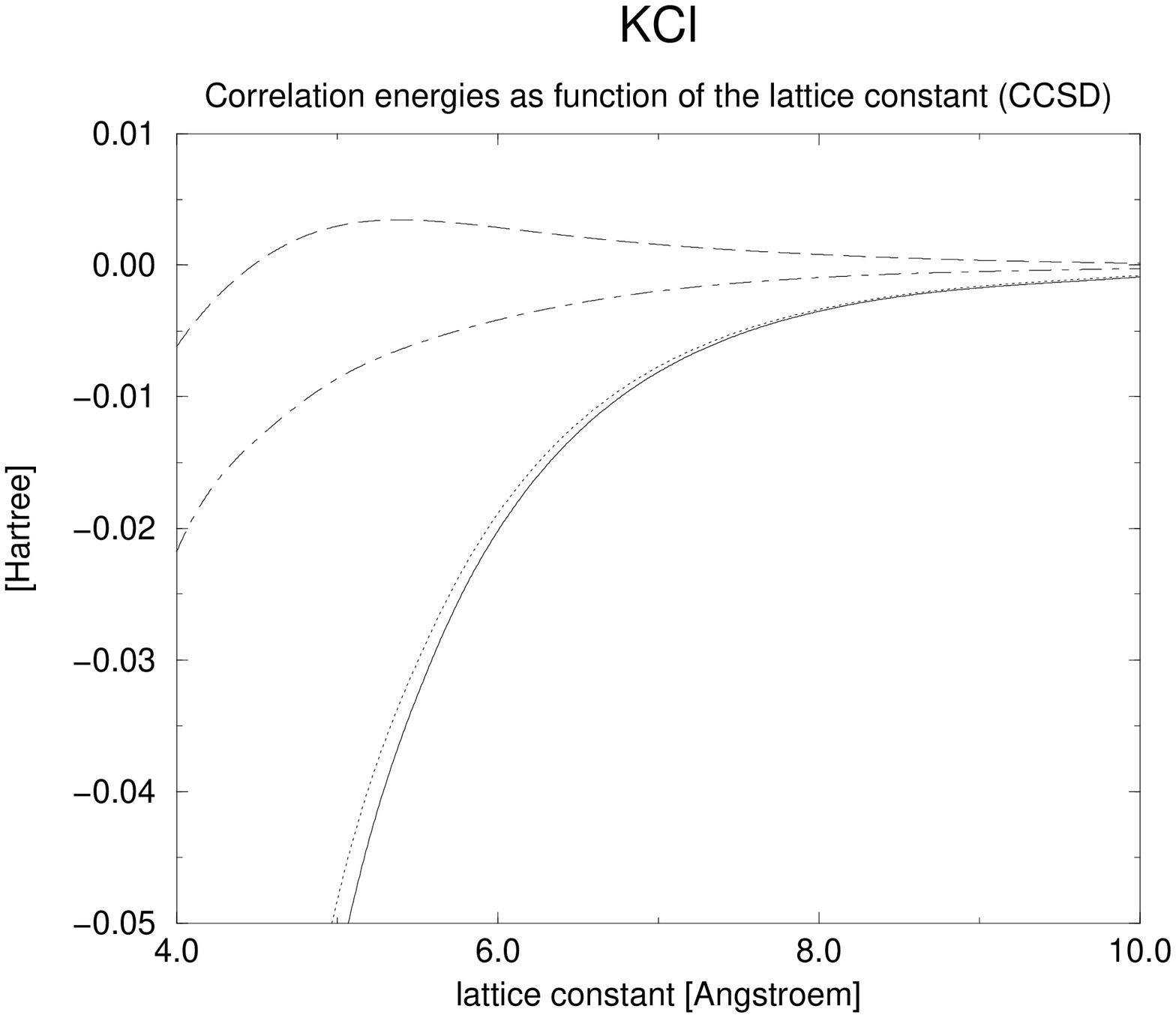,width=9cm,angle=270}}
\label{Inkrementbild}
\end{figure}

A more detailed account of correlation contributions to the 
potential-energy surface of KCl is given
in Fig. \ref{Inkrementbild}, where we display the difference of
correlation energies 
$\epsilon{\rm (embedded\mbox{ }Cl}^-)$ -$\epsilon{\rm (free\mbox{ } Cl}^-)$
as a function of the
lattice constant, i.e. its variation from
free Cl$^-$ to an embedded Cl$^-$ in KCl.
Starting from a 
very small (unrealistic) lattice constant $a$, the 
correlation energy $\epsilon{\rm (embedded\mbox{ }Cl}^-)$
decreases in magnitude with increasing $a$ -
excitations into $d_{xy}$, $d_{yz}$, $d_{xz}$-orbitals are
very important for small $a$ since these orbitals have smaller overlap 
with the region that is occupied by the K electrons -,  then passes through
a minimum and increases again because of the argument given earlier 
(excitations into the diffuse Cl $4p$ orbitals are
lower in energy the larger the distance to the K electrons).
The next-neighbor K-Cl and Cl-Cl correlation-energy increments also 
shown in Fig. \ref{Inkrementbild}  monotonously
decrease with increasing $a$, for larger distance
according to the van der Waals law.
The three contributions depicted in Fig.\ \ref{Inkrementbild} 
are the most important ones and nearly exhaust
the incremental expansion (see Table \ref{inkKCl657}, the remaining
increments amount to $\sim$ 1 mH only).
The first derivative 
of their sum with respect to the lattice constant
clearly shows that in total correlations reduce the
lattice constant. The second derivative shows that - at fixed lattice constant
- correlations reduce the bulk modulus (the one-body increments alone might
 lead to
an increase of lattice constant and bulk modulus, 
but are outweighed by the two-body increments). 

Several density functional calculations are available in literature
for the systems
considered. A
KKR calculation \cite{Yamashita} (combined with a 
local exchange-correlation potential 
\cite{GunnarssonLundqvist})  and more recently a full-potential xc-LDA
calculation \cite{Cortona} have been performed. In Refs. 
\onlinecite{CausaZupan,Apra} correlation-only density functionals with 
gradient corrections have been included
{\em a  posteriori} (i.e. using the density and non-local exchange energy
from a Hartree-Fock 
calculation). The best
density functional results are in good agreement with experiment, but it 
seems to be difficult to choose one single functional as reference method.

In Ref. \onlinecite{OchsenfeldAhlrichs}, 
a large number of alkali halide clusters has been investigated.
Bulk properties were extrapolated
from cluster calculations by linearly fitting the energy vs.\
$n^{-1/3}$, where $n$ is the number of MX units. 
The results for the lattice energies $E_{lat}$ are in good
agreement with experiment. The predicted correlation corrections
are in agreement with our findings for LiF ($\sim$ 0), but different
for NaCl (an increase of $|E_{lat}|$ of $\sim$ 0.003 H is reported, 
we find $\sim$ 0.013 H at the CCSD level) and
KCl ($\Delta |E_{lat}| \sim$ -0.011 H from Ref.\
\onlinecite{OchsenfeldAhlrichs}, we obtain a CCSD value of  $\sim$ 0.016 H). 
The geometries were optimized at the HF level using a $M_{32}N_{32}$ cluster.
It was proposed to use the bond length of the interior cube of this cluster as
an estimate of the lattice constant of the solid. This leads to a slight
underestimation in all cases compared to the HF lattice constants from 
{\sc Crystal}
calculations. Surface effects are probably the explanation for the
differences, since each atom of the interior cube has three next neighbors
also residing in the interior cube, but also three next neighbors 
located at the surface whose charges will be different from interior ions;
the Pauli repulsion and the
Madelung field are probably not too well reproduced.
This is avoided in our approach since a cluster approach is applied 
at the correlated level only, and even there all
explicitly treated ions are surrounded by pseudopotentials (or point charges)
simulating bulk cations (or anions).

\section{Conclusion}
We have shown that the method of local increments can successfully be applied 
for the determination of bulk electron-correlation effects in
alkali halides. The main shortcoming of the Hartree-Fock approximation is
the missing inter-ionic
van der Waals interaction which results in too large 
lattice constants (by up to 5\%). 
After including correlations at the coupled-cluster level, the 
deviations of the lattice constant from the experimental values
are reduced to a maximum of 1.1\%.
We obtain between 95 and 98 \% of the cohesive energies with
respect to neutral atoms or 97 to 98 \% of the lattice energies. 
Bulk moduli  exhibit satisfactory  agreement
with experiment, with a maximum deviation of $\sim$ 8\%.

\begin{table}
\begin{center}
\caption{\label{EAIP}Electron affinities and ionization potentials
in Hartrees. (1 H = 27.2114 eV) }
\vspace{5mm}
\begin{tabular}{|ccccc|}
System & HF & CCSD & CCSD(T) & exp. \cite{CRC} \\ \hline
  F $\rightarrow$ F$^{-}$ & 0.05070 & 0.11612 & 0.12192 & 0.12499 \\
  Cl $\rightarrow$ Cl$^{-}$ & 0.09505 & 0.12605 & 0.12919 & 0.13276 \\
  Li $\rightarrow$ Li$^{+}$ & 0.19631 & 0.19731 & 0.19733 & 0.19814 \\
  Na $\rightarrow$ Na$^{+}$ & 0.18195 & 0.18785 & 0.18810 & 0.18886 \\
  K  $\rightarrow$ K$^{+}$ & 0.14679 & 0.15637 & 0.15723 & 0.15952 \\
\end{tabular}
\end{center}
\end{table}

\begin{table}
\begin{center}
\caption{\label{diatomic}
Bond lengths $R_e$ (\AA), 
dissociation energies $D_e$ (eV) and vibrational frequencies
$\omega_e$ (cm$^{-1}$) of diatomic molecules. The values taken
from literature are CI(SD) calculations.}
\vspace{5mm}
\begin{tabular}{|cccccc|}
& RHF & CCSD  & CCSD(T) & literature \cite{Bauschlicher} 
& expt. \cite{Huber-Herzberg} 
\\ \hline
LiF & { } & { } & { } & { } & \\
R$_e$ & 1.555 & 1.561 & 1.565 & 1.571 & 1.564 \\
$\omega_e$ & 943 &923 & 910 & 919 & 910 \\
D$_e$ & 4.12 & 5.85 &  5.98 & 6.12 & 5.97  \\
 { } & { } & { } & { } & { } & \\
NaF & { } & { } & { } & { } & \\
R$_e$ & 1.924 & 1.925 & 1.929 & 1.921 & 1.926 \\
$\omega_e$ & 549 & 517 & 512 & 538 & 536 \\
D$_e$ & 3.11 & 4.77 &  4.91 & 5.02 & 5.36  \\
 { } & { } & { } & { } & { } & \\
KF & { } & { } & { } & { } & \\
R$_e$ & 2.204 & 2.189 & 2.189 & 2.184 & 2.171 \\
$\omega_e$ & 420 & 422 & 421 & 428  & 428 \\
D$_e$ & 3.29 & 4.89 & 5.03 & 5.10  & 5.10 \\
 { } & { } & { } & { } & { } & \\
LiCl & { } & { } & { } & { } & \\
R$_e$ & 2.037 & 2.026 & 2.028 & 2.033 & 2.021 \\
$\omega_e$ & 645 & 645 & 642 & 646 & 643 \\
D$_e$ & 3.85 & 4.76 & 4.84 & 4.86 & 4.88 \\
 { } & { } & { } & { } & { } & \\
NaCl & { } & { } & { } & { } & \\
R$_e$ & 2.390 & 2.344 & 2.344 & 2.366 & 2.361 \\
$\omega_e$ & 359 & 368 & 367 & 361 & 366 \\
D$_e$ & 3.26 & 4.12 & 4.20 & 4.23 & 4.25 \\
 { } & { } & { } & { } & { } & \\
KCl & { } & { } & { } & { }  & \\
R$_e$ & 2.738 & 2.692 & 2.688 & 2.697 & 2.667 \\
$\omega_e$ & 266 & 276 & 276 & 273 & 281 \\
D$_e$ & 3.48 & 4.21 & 4.29 & 4.33 & 4.36 \\
\end{tabular}
\end{center}
\end{table}

\onecolumn
\newpage
\begin{table}
\begin{center}
\caption{\label{alleInk}
Hartree-Fock (HF) and correlated results (CCSD, CCSD(T)), 
in comparison to density-functional (DFT) and experimental values, 
for the solids. Cohesive energies $E$ (with respect to neutral atoms) 
and lattice energies $E_{lat}$ (with respect to free ions) are
given in Hartree units, lattice constants $a$
in \AA \mbox{} and bulk moduli $B$ in GPa. Zero point energies have been 
estimated with a Debye approximation (Debye temperatures taken from Ref.
\protect{\onlinecite{Debye}}) and added to the experimental cohesive
energies. The experimental bulk moduli are at 4.2 K and 
have been taken from Ref. 
\protect{\onlinecite{Cortona}}
and references therein.}
\vspace{5mm}
\begin{tabular}{|cccccc|}
 & HF & CCSD & CCSD(T) & DFT & expt. \cite{CRC,Landolt,Cortona}
\\ \hline
LiF &  &  &  & & \\
$E_{lat}$ & 0.3975 
& 0.3976 
& 0.3961 
& 0.417$^a$, 
0.400$^b$, 
0.365$^c$ 
& 0.404 \\ 
$E$ & 0.2534 
& 0.3179 
& 0.3222 
& 0.352$^d$, 0.345 $^e$ & 0.331 \\
$a$ & 4.011 
& 3.991 
& 3.993 
& 4.035 $^d$, 4.05 $^e$
3.88$^a$, 3.96$^b$, 4.13$^c$  
& 4.010 \\
$B$ & 78.9 & 70.1 & 74.9 & 78.3 $^d$, 70.5 $^e$,
95 $^a$, 83 $^b$, 60 $^c$  
& 69.9 \\
 & & & & & \\
NaF &  &  &  &  & \\ 
$E_{lat}$ & 0.3496
& 0.3518
& 0.3504
&  & 0.358 \\
$E$ & 0.2186 & 0.2803 & 0.2845 & 0.323 
$^d$, 0.294 $^e$ & 0.294 \\
$a$ & 4.636
& 4.601  
& 4.603
& 4.582 $^d$, 4.76 $^e$ & 4.609 \\
$B$ & 52.2 & 55.7 & 53.9
& 55.8 $^d$, 42.3 $^e$ & 51.4 \\
& & & & & \\
KF &  &  &  &  & \\ 
$E_{lat}$ & 0.3028 & 0.3101 & 0.3100  & & 0.318 \\
$E$ & 0.2076 
& 0.2707 
& 0.2755 
& 0.294 $^e$ & 0.283 \\
$a$ & 5.450 
& 5.331 
& 5.320 
& 5.40 $^e$ & 5.311 \\
$B$ & 29.9 
& 34.4 
& 34.8 
& 31.3 $^e$ & 34.2 \\
& & & & & \\
LiCl &  &  &  &  & \\ 
$E_{lat}$ & 0.3088  & 0.3225 & 0.3241 & & 0.331 \\
$E$ & 0.2096 & 0.2533 & 0.2580 
& 0.251 $^d$, 0.265 $^e$  & 0.266 \\
$a$ & 5.281 & 5.136 & 5.124 
& 5.32 $^d$, 5.08 $^e$ & 5.106 \\
$B$ & 30.1  & 35.2  & 34.8 
& 28 $^d$, 35.2 $^e$ & 35.4 \\
& & & & & \\
NaCl &  &  &  &  &  \\
$E_{lat}$ & 0.2839 & 0.2960 & 0.2971 
& 0.304$^a$, 0.312$^b$, 0.285$^c$, 
0.307$^f$, 0.303$^g$, 0.300$^{h,i}$  
& 0.302 \\ 
$E$ & 0.1978 
& 0.2350 & 0.2390 & 0.239$^d$, 0.232$^{e,i}$ & 0.246 \\
$a$ & 5.791 & 5.646 & 5.634 
& 5.737 $^d$, 5.75 $^e$,
5.47$^a$, 5.49$^b$,  5.83$^c$,
5.53$^f$, 5.51$^g$, 5.54$^{h,i}$  
& 5.595 \\
$B$ & 24.5 
& 26.6 
& 26.6 
& 25.5 $^d$, 22.8  $^e$,
31$^a$, 29$^b$, 21$^c$,
32.5$^f$, 32.1$^g$, 30.1$^{h,i}$  & 26.6 \\
& & & & & \\
KCl &  &  &  &  & \\
$E_{lat}$ & 0.2538 & 0.2687 & 0.2704 &  & 0.275 \\ 
$E$ & 0.2035 & 0.2398 & 0.2438 
& 0.249 $^d$, 0.243 $^e$ & 0.248 \\
$a$ & 6.548 & 6.314 & 6.295 
& 6.30 $^d$, 6.26 $^e$ & 6.248 \\
$B$ & 15.5 & 18.4 & 21.3 & 19.7 $^d$, 18.9 $^e$
& 19.7 \\
\end{tabular}
\end{center}
\end{table}
\mbox{ }\\
$^a${Ref. \onlinecite{CausaZupan}, Hartree-Fock exchange, Perdew and Wang
91 correlation functional \cite{PerdewWang}}\\
$^b${Ref. \onlinecite{CausaZupan}, LDA exchange and correlation}\\
$^c${Ref. \onlinecite{CausaZupan}, Becke exchange \cite{Becke} and 
Perdew and Wang 91 correlation functional \cite{PerdewWang}}\\
$^d${Ref. \onlinecite{Yamashita}, KKR calculation with local exchange and
correlation \cite{GunnarssonLundqvist}\\
$^e${Ref. \onlinecite{Cortona}, LDA exchange and correlation}\\
$^f${Ref. \onlinecite{Apra}, Hartree-Fock exchange, 
Colle and Salvetti correlation  functional
\cite{ColleSalvetti}}\\
$^g${Ref. \onlinecite{Apra}, Hartree-Fock exchange, 
Perdew 1986 correlation functional}
\cite{P86} \\
$^h${Ref. \onlinecite{Apra}, Hartree-Fock exchange, Perdew and Wang
91 correlation  functional \cite{PerdewWang}}\\
$^i${For further density functional results for NaCl, see also Ref. 
\onlinecite{Apra}}
\twocolumn

\onecolumn
\begin{table}
\begin{center}
\caption{\label{inkfluorides}Local correlation-energies 
per primitive unit cell (in Hartree)
for LiF ( at a lattice constant of 3.99 \AA), NaF (4.60 \AA), and KF
(5.34 \AA)}
\vspace{5mm}
\begin{tabular}{|c|cc|cc|cc|} 
 & \multicolumn{2}{c|}{LiF} & \multicolumn{2}{c|}{NaF} 
& \multicolumn{2}{c|}{KF} \\
 & \multicolumn{1}{c}{CCSD}  & \multicolumn{1}{c|}{CCSD(T)} 
 & \multicolumn{1}{c}{CCSD}  & \multicolumn{1}{c|}{CCSD(T)} 
 & \multicolumn{1}{c}{CCSD}  & \multicolumn{1}{c|}{CCSD(T)} \\
\hline
 {free X$^{+}$} $\rightarrow$ \rm {embedded X$^{+}$} & -0.000021 &
 -0.000021  & -0.000165$^a$ & -0.000179$^a$ 
& -0.000013 & -0.000016 \\
 {free F$^{-}$}  $\rightarrow$ \rm {embedded F$^{-}$} &   +0.011776 & 
+0.014782 & +0.010106 & +0.012820 &   +0.009684 & +0.012352 \\
 {sum of F-F increments} & -0.007926 & -0.009141 & -0.003384 & -0.003954 
& -0.001170 & -0.001359  \\
 {sum of X-F increments}  & -0.003970 & -0.004254 & -0.008554 & -0.009346
 & -0.014940 & -0.017074 \\ 
 {sum of X-X increments}   & -0.000018 & -0.000018 & -0.000198 & -0.000210 
& -0.001422 & -0.001605 \\ 
 & & & & & & \\
 {sum  } & -0.000159 & +0.001348 
& -0.002195 & -0.000869  & -0.007861 & -0.007702\\
\end{tabular}
\end{center}
\end{table}
$^a$ See footnote (Ref. \onlinecite{Kontraktion}).

\begin{table}
\begin{center}
\caption{\label{inkchlorides}Local 
correlation-energies per primitive unit cell (in Hartree)
for LiCl (at a lattice constant of 5.14 \AA), NaCl (5.65 \AA), and KCl 
(6.30 \AA).}
\vspace{5mm}
\begin{tabular}{|c|cc|cc|cc|} 
 & \multicolumn{2}{c|}{LiCl} & \multicolumn{2}{c|}{NaCl}
& \multicolumn{2}{c|}{KCl} \\
 & \multicolumn{1}{c}{CCSD}  & \multicolumn{1}{c|}{CCSD(T)} 
 & \multicolumn{1}{c}{CCSD}  & \multicolumn{1}{c|}{CCSD(T)} 
 & \multicolumn{1}{c}{CCSD}  & \multicolumn{1}{c|}{CCSD(T)} \\
\hline
 {free X$^{+}$} $\rightarrow$ \rm {embedded X$^{+}$} & -0.000013 &
 -0.000013 & -0.000101$^a$ & -0.000109$^a$ & -0.000005 & -0.000005 \\
 {free Cl$^{-}$}  $\rightarrow$ \rm {embedded Cl$^{-}$} &   +0.002567 & 
+0.003572 &   +0.002448 & +0.003415 &   +0.002426 & +0.003411\\
 {sum of Cl-Cl increments} & -0.014439 & -0.016785 & -0.008124 & -0.009495 
 & -0.003732 & -0.004368 \\
 {sum of X-Cl increments}  & -0.002712 & -0.002906 & -0.007112 & -0.007746 
& -0.014992 & -0.017066 \\ 
 {sum of X-X increments}   & \multicolumn{2}{c|}{absolute value $<$10$^{-6}$}
 & -0.000054 & -0.000060 & -0.000444 & -0.000501  \\
 {sum of three-body increments} & - & - & - & - &  +0.000388 & +0.000372 \\
{ } & & & & & & \\ 
{sum } & -0.014597 & -0.016132 & -0.012943 & -0.013995
&  -0.016359 & -0.018157 \\
\end{tabular}
\end{center}
\end{table}
$^a$ See footnote (Ref. \onlinecite{Kontraktion}).
\twocolumn

\begin{table}
\begin{center}
\caption{\label{inkKCl657}Local
correlation-energies per primitive unit cell (in Hartree)
 for KCl at a lattice constant of 
6.57 \AA. The quantities involving 2 and 3 ions are non-additivity 
corrections (increments).}
\vspace{5mm}
\begin{tabular}{|c|c|c|c|} 
& weight & \multicolumn{1}{c|}{CCSD}  & \multicolumn{1}{c|}{CCSD(T)} \\
\hline
 {free K$^{+}$} $\rightarrow$ \rm {embedded K$^{+}$} & 1 & -0.000004 &
 -0.000004 \\
 {free Cl$^{-}$}  $\rightarrow$ \rm {embedded Cl$^{-}$} & 1 &  +0.002059 & 
+0.002911 \\
\rm{Cl(0,0,0)-Cl(0,1,1)} & 6 & -0.002736 & -0.003228 \\
\rm{Cl(0,0,0)-Cl(2,0,0)} & 3 & -0.000138 & -0.000162 \\
\rm{Cl(0,0,0)-Cl(2,1,1)} & 12 & -0.000144 & -0.000168 \\
\rm{Cl(0,0,0)-Cl(2,2,0)} & 6 & -0.000030 & -0.000036 \\
\rm{K(0,0,0)-Cl(1,0,0)} & 6 & -0.011256 & -0.012858 \\
\rm{K(0,0,0)-Cl(1,1,1)} & 8 & -0.000320 & -0.000360 \\
\rm{K(0,0,0)-Cl(2,1,0)} & 24 & -0.000192 & -0.000216 \\
\rm{K(0,0,0)-K(0,1,1)} & 6 & -0.000318 & -0.000360 \\
\rm{K(0,0,0)-K(2,0,0)} & 3 & -0.000018 & -0.000021 \\
 {Cl(1,0,0)-Cl(0,1,0)-Cl(0,0,1)} & 8 & +0.000064 & +0.000080 \\
 {Cl(0,0,0)-K(0,1,0)-Cl(0,1,1)} & 12 & +0.000204 & +0.000204 \\
{ } & & & \\  
{sum } & & -0.012829 & -0.014218 \\
\end{tabular}
\end{center}
\end{table}

\onecolumn
\begin{table}
\begin{center}
\caption{\label{vdWTabelle}Comparison of CCSD two-body increments 
$\Delta E$
between next
neighbors (without multiplying with the weight factor). All results are
given in atomic units (except for the lattice constant in column 2). 
$r$ is the distance
between the respective ions in bohr.}
\vspace{5mm}
\begin{tabular}{|ccccccccc|} 
 System & lattice & \multicolumn{1}{c}{$\Delta E$}  &
 $\Delta E$ $\times$ $r^6$ & IP$_{cat}$ & IP$_{an}$ &
$\alpha_{cat}$ & $\alpha_{an}$ & $ -\frac{2}{3}\frac{r^6}{\alpha_{1}
\alpha_{2}}\frac{IP_1 + IP_2}{IP_1 IP_2}$ $\times$ $\Delta E$\\
& constant $a$ in \AA &  & & & & & & \\
\hline
F-F (LiF) & 3.99 & -0.001181  & -27.1 & 2.3 & 0.52 & 0.19 & 5.0 & 2.8 \\
F-F (NaF) & 4.60 & -0.000502 & -27.1 & 1.3 & 0.47 & 0.97 & 5.4 & 2.6 \\
F-F (KF) & 5.34 & -0.000174 &  -23.0 & 0.80 & 0.42 & 5.4 & 5.4 & 2.5 \\
Cl-Cl (LiCl) & 5.14 & -0.002155  & -226 & 2.4 & 0.45 & 0.19 & 19 & 1.9 \\
Cl-Cl (NaCl & 5.65 & -0.001215  & -225 & 1.4 & 0.42 &  0.97 & 19 & 2.0 \\
Cl-Cl (KCl) & 6.30 & -0.000558  & -199 & 0.85 & 0.39 & 5.4 & 18 & 2.1 \\
O-O (MgO)$^a$ & 4.18 & -0.002582  & -78.4 & 2.1 & 0.38 & 0.48 & 9.7 & 2.9 \\
O-O (CaO)$^a$ & 4.81 & -0.001067  & -75.2 & 1.2 & 0.27 & 3.1  & 9.7 & 3.9 \\
O-O (NiO)$^a$ & 4.17 & -0.003356  & -100 & 0.42 & 0.41 & 2.8 & 11.4 & 2.5 \\
Li-F & 3.99 & -0.000627  & -1.80 &  &   &  &  & 3.0\\
Na-F & 4.60 & -0.001351  & -9.11 & & & & & 3.4\\
K-F & 5.34 & -0.002382  & -39.3 & & & & & 3.3\\
Li-Cl & 5.14 & -0.000440  & -5.77 & & & & & 2.8\\
Na-Cl & 5.65 & -0.001132  & -26.2 & & & & & 2.9\\
K-Cl & 6.30 & -0.002392  & -106 & & & & & 2.7 \\
Mg-O$^a$ & 4.18 & -0.003129  & -11.9 & &  &  &  & 5.3 \\
Ca-O$^a$ & 4.81 & -0.005906  & -52.0 & & & & & 5.2 \\
Ni-O$^a$ & 4.17 & -0.009958 & -37.3 & & & & & 3.8 \\
\end{tabular}
\end{center}
\end{table}
$^a$ Ref. \onlinecite{DDFS}
\twocolumn


\begin{references}
\bibitem{KBasis} {5 even tempered $d$-exponents  }
(3.285600, 0.888000, 0.240000, 0.064865, 0.017531) were optimized 
in CCSD calculations for the $^2D$
state and in addition 3 even tempered
$f$-functions (exponents 2.530000,1.100000,0.478261) for the $^2S$ state.
\bibitem{Kalium} For KF, we optimized one $d$-exponent for K (0.54), for KCl we
optimized one $d$-exponent for K (0.44) and one $d$ exponent 
for Cl (0.5, a more diffuse exponent led to numerical instabilities). 
The misprints of the original paper (Ref. \onlinecite{Prencipe}; 
+0.1222 instead of -0.1222 
in the F basis set, and
+0.0082 instead of -0.082 in the Cl basis set) have been corrected.
To calculate the energies of the neutral Li, Na, and K atoms,
enhanced basis sets have to be used, we enlarged the basis sets by
 one more $sp$ exponent.
\bibitem{CCSDT}As already pointed out in Ref. \onlinecite{DDFS}, the triples
correction within CCSD(T) is only exact for canonical orbitals. Since we
have to used localized orbitals, we estimated
the error as described in previous work and did not find it to be a
significant error.
\bibitem{Kontraktion} The difference in correlation energies 
free Na$^+$ $\rightarrow$ embedded Na$^+$
is relatively large compared to Li and K. This changes when the Na basis set
is taken uncontracted. Of course, the error for the total energy is 
negligibly small.
We did not find any other artifact of the contraction.
\end{references}
\end{document}